\documentclass[aps,prb,twocolumn,showpacs,amsfonts]{revtex4-1}
\usepackage{epsfig}\usepackage{subfigure}
\usepackage{graphicx,graphics}
\usepackage{amsmath,ulem}
\usepackage{amssymb}
\usepackage{bm}
\usepackage{color}
\usepackage[utf8]{inputenc}

\begin{document}

\title{Proximity-Induced Superconductivity at Non-Helical Topological Insulator Interfaces}

\author{David J. Alspaugh}
\affiliation{Department of Physics and Astronomy, Louisiana State University, Baton Rouge, LA 70803-4001}

\author{Mahmoud M. Asmar}
\affiliation{Department of Physics and Astronomy, Louisiana State University, Baton Rouge, LA 70803-4001}

\author{Daniel E. Sheehy}
\affiliation{Department of Physics and Astronomy, Louisiana State University, Baton Rouge, LA 70803-4001}

\author{Ilya Vekhter}
\affiliation{Department of Physics and Astronomy, Louisiana State University, Baton Rouge, LA 70803-4001}

\date{July 12, 2018}

\begin{abstract}
{We study how non-helical spin textures at the boundary between a topological insulator (TI) and a superconductor (SC) affect the proximity-induced superconductivity of the TI interface state. We consider TIs coupled to both spin-singlet and spin-triplet SCs, and show that for the spin-triplet parent SCs  the resulting order parameter induced onto the interface state sensitively depends on the symmetries which are broken at the TI-SC boundary. 
For chiral spin-triplet parent SCs, we find that nodal proximity-induced superconductivity emerges when there is broken twofold rotational symmetry which forces the spins of the non-helical topological states to tilt away from the interface plane. We furthermore show that the Andreev conductance of lateral heterostructures joining TI-vacuum and TI-SC interfaces yields experimental signatures of the reduced symmetries of the interface states.}
\end{abstract}


\maketitle

\textit{Introduction.}  Topological insulators (TIs) are a class of materials which belong to a distinct phase separate from their trivial counterparts despite respecting the same global symmetries\cite{Hasan2010,Qi2011}. The main signature of this phase is the presence of linearly dispersing metallic states at the boundaries of the TI. These states are robust under perturbations that preserve time reversal symmetry, and as a consequence of the bulk spin-orbit interaction their spin and momenta are locked relative to each other. 

At planar TI-vacuum terminations these gapless surface states have isotropic dispersions, are perfectly helical (with their spin normal to the direction of their momentum and confined in the interface plane), and can be described by a two-dimensional massless-Dirac effective Hamiltonian. It is commonly assumed that interfaces within heterostructures of TIs and topologically trivial materials exhibit these same properties. However, effects due to lattice strain, charge redistribution, dangling bonds, and other non-magnetic interface potentials may lower the symmetry of the interface relative to the bulk. It was recently shown that metallic states at the interface reflect these reduced symmetries, so that generally  their dispersion is anisotropic and the spin-momentum locking is not helical\cite{Asmar2017}.

Some of the most promising potential applications of TIs rely on the proximity-induced superconductivity from a TI interface state in contact with a bulk superconductor (SC)\cite{Fu2008}. When vortices are present or when placed alongside ferromagnetic systems, these junctions are predicted to host Majorana fermions, which are critical for fault-tolerant quantum computing~\cite{Fu2008,Read2000}. However, the properties of the induced superconductivity strongly depend on the spin structure of the interface state as well as the properties of the parent SC.  Existing conclusions about the interface superconductivity, including the prediction that some spin-triplet parent SCs do not induce superconductivity in the interface state at all~\cite{Linder2010a}, have been reached assuming helical TI surface states.  

\begin{figure}[t]
\includegraphics[width=\columnwidth]{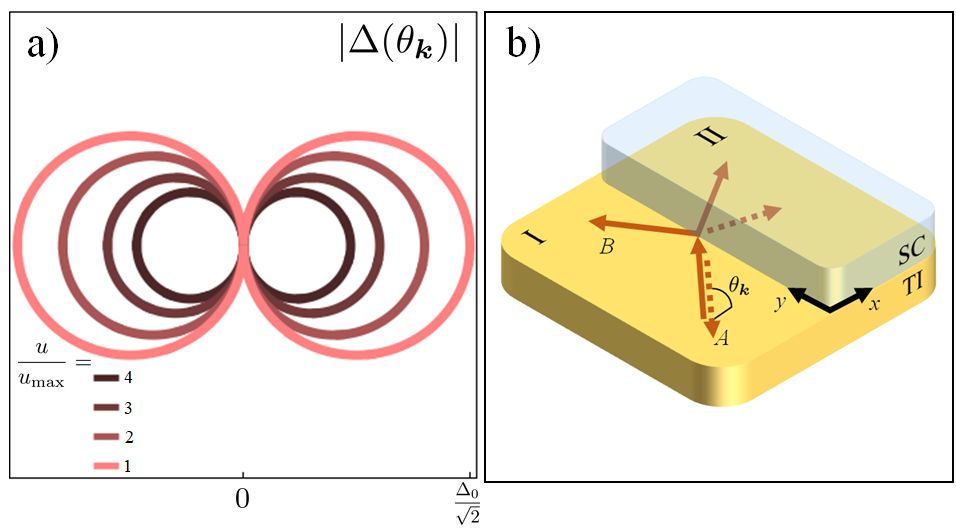}
\caption{Essential aspects of the proximity pairing in the proposed experimental setup. Panel a) Polar plot of the induced superconducting gap magnitude as a function of the in-plane momentum angle $\theta_{\bm{k}} = \tan^{-1}k_{y}/k_{x}$ for an interface with a chiral spin-triplet parent superconductor. Legend gives the strength of the rotational symmetry breaking interface potential as modeled by Eq.~\eqref{mahmoud}. Panel b) Schematic of the lateral heterojunction. In region I, the TI surface state is described by $H_{\rm D}(\bm{k}) = \hbar v_{\rm F}(\bm{\sigma}\times\bm{k})_{z}$. In region II, the TI interface state is is non-helical. Arrows indicate the quasiparticle states participating in the Andreev reflection process.}
\label{fig:mags}
\end{figure}

In this work we analyze  proximity-induced superconductivity in TI-SC heterostructures assuming the most general form of the interface state allowed by symmetry. We obtain the superconducting order parameter at the interface and demonstrate that when the parent SC is spin-singlet, the shape of the induced superconducting gap mimics that of the parent. For  spin-triplet parent SCs on the other hand, distinct anisotropic phases emerge depending on the nature of the interface. Strikingly, interface potentials may enable proximity-induced superconductivity for spin-triplet SCs that do not induce superconductivity in the helical Dirac states. One such example is the chiral state suggested for Sr$_2$RuO$_4$~\cite{Mackenzie2003}, which requires out-of-plane spin textures in the TI interface state. In addition,
we show that conductance spectroscopy of lateral heterostructures of TI-vacuum and TI-SC interfaces exhibits 
clear signatures of symmetry-breaking interface potentials.

\textit{Model for the interface states.} To describe TI interface states, we start from the most general time-reversal invariant Hamiltonian that is linear in momentum:
\begin{equation}
H(\bm{k}) =  \bm{c}(\bm{k})\cdot\bm{\sigma}\,.
\label{tiham}
\end{equation}
Here, $\bm{k} = (k_{x},k_{y})$ is the in-plane momentum, $\bm{\sigma} = (\sigma_{x},\sigma_{y},\sigma_{z})$ is a vector of  Pauli matrices in spin space, and $\bm c(\bm k)$ is a three-dimensional vector such that $c_{i}(\bm{k}) = \sum_{j} c_{ij}k_{j}$ with real coefficients $c_{ij}$, where $i\in \{x,y,z\}$ and $j\in \{x,y\}$. We contrast the general Hamiltonian Eq.~(\ref{tiham}) with the well-known Dirac Hamiltonian describing the low-energy physics of TI surface states, $H_{\rm D}(\bm{k}) = \hbar v_{\rm F}(\bm{\sigma}\times\bm{k})_{z}$, which is obtained for the choice $\bm{c}_{\rm D}(\bm{k}) = \hbar v_{\rm F}(k_{y},-k_{x},0)$ where $v_{\rm F}$ is the Fermi velocity. 

Due to the mismatch of the basis functions it is generally difficult to develop reliable effective models for SC-semiconductor heterostructures~\cite{Mikkelsen2018}, and especially so for topological systems~\cite{Zhao2010}. While $H_{\rm D}(\bm{k})$ is often believed to hold for idealized interfaces of TIs with non-topological materials, Ref.~\onlinecite{Asmar2017} showed that non-magnetic interface potentials generally lead to
an interface described by Eq.~(\ref{tiham}) with modified coefficients $c_{ij}$, whose values are determined by the material-specific details in microscopic calculations. The general form of these coefficients however can be determined by imposing spatial symmetries at the interface, and therefore we use Eq.~\eqref{tiham} as a general starting point to analyze the proximity effect in TI-SC heterostructures. The description of the TI interface states via Eq.~(\ref{tiham}) is valid as long as the the coupling between the TI and the SC is small, so that the interface states retain their topological character~\cite{Grein2012,Zhao2010}.

\textit{Interface Superconductivity.}  Since $\bm c(\bm k)=-\bm c(-\bm k)$, the form of the Hamiltonian in Eq.~\eqref{tiham} is identical to that of antisymmetric spin-orbit coupling in non-centrosymmetric metals, whose influence on superconducting pairing has been extensively studied~\cite{Smidman2017}. The key difference is that the regular quadratic kinetic energy term is absent, placing us in the limit of infinitely strong spin-orbit coupling \cite{Pesin2012}. Consequently, only one of the spin-orbit split bands crosses the chemical potential, and the quasiparticles that form Cooper pairs at the TI interface are effectively spinless.

In this limit, and under the assumption of weak TI-SC coupling, the proximity-induced superconducting order in the interface layer can be adequately described by simply projecting the Cooper pair structure of the parent SC, $\hat{\Delta} = [\psi(\bm{k}) + \bm{d}(\bm{k})\cdot\bm{\sigma}](i\sigma_{y})$ where $\psi(\bm{k})$ and $\bm{d}(\bm{k})$ are the spin-singlet and spin-triplet parts of the pairing field respectively,  onto the eigenstates of the interface Hamiltonian. Writing the electron field operators as $(a_{\bm{k}\uparrow},a_{\bm{k}\downarrow})$, and introducing the Nambu spinor $\Psi = (a_{\bm{k}\uparrow},a_{\bm{k}\downarrow},a_{-\bm{k}\uparrow}^{\dagger},a_{-\bm{k}\downarrow}^{\dagger})$, the Bogoliubov-de Gennes (BdG) Hamiltonian becomes
\begin{equation}
\mathcal{H} = \dfrac{1}{2}\sum_{\bm{k}}\Psi^{\dagger}\begin{pmatrix}
H(\bm{k})-\mu & \hat{\Delta}
\\ \hat{\Delta}^{\dagger} & -H^{T}(-\bm{k})+\mu
\end{pmatrix}
\Psi.
\label{bdg}
\end{equation}
Here $\mu$ is the chemical potential. Transforming to the band representation, denoting  the annihilation operators for the eigenstates of the conduction and the valence bands in Eq.~\eqref{tiham} as $(b_{\bm{k}1},b_{\bm{k}2})$ respectively, and introducing the corresponding band Nambu spinor $\Phi = (b_{\bm{k}1},b_{\bm{k}2},b_{-\bm{k}1}^{\dagger},b_{-\bm{k}2}^{\dagger})$, we arrive at the mean field pairing Hamiltonian for the interface,
\begin{widetext}
\begin{equation}
\mathcal{H} = \dfrac{1}{2}\sum_{\bm{k}}\Phi^{\dagger}    \begin{pmatrix}
|\bm{c}| - \mu & 0 & -e^{-i\varphi_{\bm{c}}}[\hat{\bm{c}}\cdot \bm{d} + \psi] & \tfrac{[\hat{\bm{c}}\times(\hat{\bm{c}}\times\bm{d})]_{z} - i(\hat{\bm{c}}\times\bm{d})_{z}}{e^{i\varphi_{\bm{c}}} \sin\vartheta_{\bm{c}}} 
\\ 0 & -|\bm{c}| - \mu &  -\tfrac{[\hat{\bm{c}}\times(\hat{\bm{c}}\times\bm{d})]_{z} + i(\hat{\bm{c}}\times\bm{d})_{z}}{e^{i\varphi_{\bm{c}}} \sin\vartheta_{\bm{c}}}   & -e^{-i\varphi_{\bm{c}}}[\hat{\bm{c}}\cdot \bm{d} - \psi]
\\ -e^{i\varphi_{\bm{c}}}[\hat{\bm{c}}\cdot \bm{d}^{*} + \psi^{*}] &  -\tfrac{[\hat{\bm{c}}\times(\hat{\bm{c}}\times\bm{d}^{*})]_{z} - i(\hat{\bm{c}}\times\bm{d}^{*})_{z}}{e^{-i\varphi_{\bm{c}}}\sin\vartheta_{\bm{c}} } & -|\bm{c}| + \mu & 0
\\ \tfrac{[\hat{\bm{c}}\times(\hat{\bm{c}}\times\bm{d}^{*})]_{z} +i(\hat{\bm{c}}\times\bm{d}^{*})_{z}}{e^{-i\varphi_{\bm{c}}}\sin\vartheta_{\bm{c}}} & -e^{i\varphi_{\bm{c}}}[\hat{\bm{c}}\cdot \bm{d}^{*} - \psi^{*}] & 0 &  |\bm{c}| + \mu
\end{pmatrix}  \Phi.
\label{ham}
\end{equation}
\end{widetext}
Here $\hat{\bm{c}}(\bm{k})$ is the unit vector along $\bm{c}(\bm{k})$ and $\vartheta_{\bm{c}}(\bm{k}),\varphi_{\bm{c}}(\bm{k})$ are the polar and azimuthal angles of the vector $\bm{c}(\bm{k})$, respectively. For brevity in Eq.~\eqref{ham} we omitted the argument $\bm k$, but the momentum dependence is implicit in all of the functions. In principle, Eq.~\eqref{ham} contains all possible superconducting order parameters allowed in a TI-SC heterostructure, including both the intra- and inter-band pairing of the interface states. Below we focus on the doped case in the weak-coupling limit $\mu\gg |\psi(\bm k)|, |\bm d(\bm k)|$, when the interband pairing can be ignored and from Eq.~(\ref{ham}) we can see that the order parameter for the  spinless fermions in the conduction band takes the form 
\begin{equation}
\Delta(\bm{k}) = -e^{-i\varphi_{\bm{c}}(\bm{k})}\Big(\hat{\bm{c}}(\bm{k})\cdot \bm{d}(\bm{k}) + \psi(\bm{k})\Big).
\label{op}
\end{equation}
The implications of Eq.~(\ref{op}) are distinct for the cases of singlet and triplet SCs.  In the singlet case, the momentum dependence of the gap magnitude $|\Delta(\bm{k})|$ is the same as that of the 
parent bulk SC. However, the phase winding of the order parameter yields nontrivial phenomena. While the connection of $\varphi_{\bm{c}}(\bm{k})$ to the angle of the in-plane momentum $\theta_{\bm{k}} = \tan^{-1}k_{y}/k_{x}$ depends on the $c_{ij}$ coefficients, the condition $c_{xx}c_{yy}\neq c_{xy}c_{yx}$ guarantees that $\varphi_{\bm{c}}(\bm{k})$ always winds by $2\pi$ along any closed path around the $\Gamma$-point of the surface Brillouin zone. This is analogous to the case studied in Ref. \onlinecite{Fu2008} where Majorana fermions have been demonstrated to appear within TI-SC heterostructures involving helical TI surface states.

In the triplet case Eq.~(\ref{op}) implies that the proximity-induced pairing is
sensitive to the detailed form of the interface Hamiltonian, and the shape of the proximity induced gap generally differs from that of the parent SC. Even for a fully gapped bulk SC, it is possible to have nodal interface superconductivity. This is evident already for $H_{\rm D}(\bm{k})$, where we obtain $\Delta_{\rm D}(\bm{k}) = -ie^{-i\theta_{\bm{k}}}(\bm{d}(\bm{k})\times \hat{\bm{k}})_{z}$. For concreteness below we assume a tetragonal crystal symmetry for the parent superconducting material and classify $\bm{d}(\bm{k})$ according to irreducible representations of the D$_{4h}$ point group. For example, the fully gapped helical A$_{2\rm u}$ parent state $\bm d_{\rm A_{2\rm u}}(\bm k) = \Delta_{0}(\hat{k}_y \hat{\bm{x}} - \hat{k}_x \hat{\bm{y}})$ leads to  fully-gapped superconductivity in the TI layer, while the parent $\rm B_{1\rm u}$ state $\bm{d}_{\rm B_{1\rm u}}(\bm{k}) = \Delta_{0}(\hat{k}_{x}\hat{\bm{x}} - \hat{k}_{y}\hat{\bm{y}})$ and the parent $\rm B_{2\rm u}$ state $\bm d_{\rm B_{2\rm u}}(\bm k) = \Delta_{0}(\hat{k}_y \hat{\bm{x}} +\hat{k}_x \hat{\bm{y}})$ produce d-wave like nodal gaps. This result is in agreement with the symmetry-based analysis of Ref.~\onlinecite{Black-Schaffer2013}. Deviations from the vector $\bm c_{\rm D}(\bm k)$ introduce additional anisotropies into the proximity-induced gap. These reflect the lower symmetry due to interface potentials, but do not qualitatively change the gap structure.  

Qualitative differences appear for triplet parent materials with $\rm A_{1\rm u}$ or $\rm E_{2\rm u}^{\pm}$ pairing states,
$\bm d_{\rm A_{1\rm u}}(\bm k) = \Delta_{0}(\hat{k}_x \hat{\bm{x}} + \hat{k}_y \hat{\bm{y}})$ and
$\bm d_{\rm E_{2\rm u}^{\pm}}(\bm k) = \Delta_{0}(\hat{k}_x  \pm i \hat{k}_y) \hat{\bm z}$, respectively. In both cases a conventional Dirac interface Hamiltonian yields no proximity-induced superconductivity~\cite{Linder2010a,Black-Schaffer2013,Linder2010}. In contrast, we find that a more general $\bm c(\bm k)$ reflecting the effects of the interface potentials enables pairing of the interface states. For the $\rm A_{1\rm u}$ SC the induced superconducting order parameter may be nodal or fully gapped depending on the specific choice of $c_{ij}$. Below, motivated in part by the studies of SrRu$_2$O$_4$, we focus on the chiral $\rm E_{2\rm u}^{+}$ case and investigate signatures of symmetry breaking at the TI-SC interface in the proximity-induced superconductivity.

\textit{Interface Potentials and Proximity Effect with Chiral SCs.}
As is evident from Eq.~\eqref{op}, such a proximity effect requires $c_z(\bm k)\neq 0$, leading to 
an induced gap that has nodes along the directions $c_{zx}k_x+c_{zy}k_y=0$ in the interface plane.  Although at vacuum termination $H_{\rm D}(\bm{k})$
has $c_z(\bm k) =0$, this coefficient is nonzero in the presence of interface scattering 
that breaks rotational symmetry in the plane of the interface~\cite{Asmar2017}.

To proceed, we need a specific model for the $c_{ij}$ coefficients. Here we adopt results from the microscopic model of Ref.~\onlinecite{Asmar2017}, which shows that if $u$ is the strength of a symmetry-breaking interface potential at the $z = 0$ interface, then the effective interface Hamiltonian is characterized by the coefficients
\begin{equation}
\bm{c}_{u}(\bm{k}) = \hbar v_{F}\bigg(k_{y} ,-\dfrac{k_{x}}{1 + G(u)},-\dfrac{\sqrt{2G(u)}}{1 + G(u)}k_{x}\bigg)\,.
\label{mahmoud}
\end{equation}
Here $G(u) = (u^{2}/G_{0})/[1 + (u^{2}/G_{0})^{2}]\leq 1$, where $G_{0}$ is a real parameter whose numerical value is determined by the material parameters of the heterostructure.  This form for the interface Hamiltonian indeed breaks rotational symmetry, lacking reflection symmetry in the $x-z$ plane, with
the $c_z(\bm{k})\neq 0$ term generating a rotation of the interface state's spin out of the plane of the interface.

We thus expect that any interface scattering that breaks mirror symmetries will generally lead to an 
interface Hamiltonian characterized by Eq.~(\ref{mahmoud}), and now investigate its experimental
implications. In the absence of interface potentials ($u=0$) and for a strong barrier ($u^2\gg G_0$), this expression recovers the helical Dirac case $\bm c_{\rm D}(\bm k)$. The maximal deviation from $\bm c_{\rm D}(\bm k)$, where we expect the strongest proximity effect, occurs at $u_{\text{max}} = \sqrt{G_{0}}$. In Fig.~\ref{fig:mags}(a)
we show the evolution of the proximity-induced order parameter for various values of the interface potential 
strength $u$. Since $G(u)$ is an algebraic function of $u$ the eigenstates of the interface Hamiltonian, Eq.~\eqref{tiham}, retain a significant out-of-plane spin component and therefore allow for the proximity coupling to a chiral triplet, even for potential strengths far from $u_{\text{max}}$.

Note that the constant energy surfaces for the interface Hamiltonian Eq.~\eqref{tiham} with the choice of the coefficients in Eq.~\eqref{mahmoud} are elliptical, elongated in the $k_x$ direction. The complementary choice of a symmetry breaking interface potential gives the ellipse elongated along the $k_y$ axis. In general, however, the existence of $c_z(\bm k)$ need not require ellipticity of the Fermi surface. 

\begin{figure}
\includegraphics[width=\columnwidth]{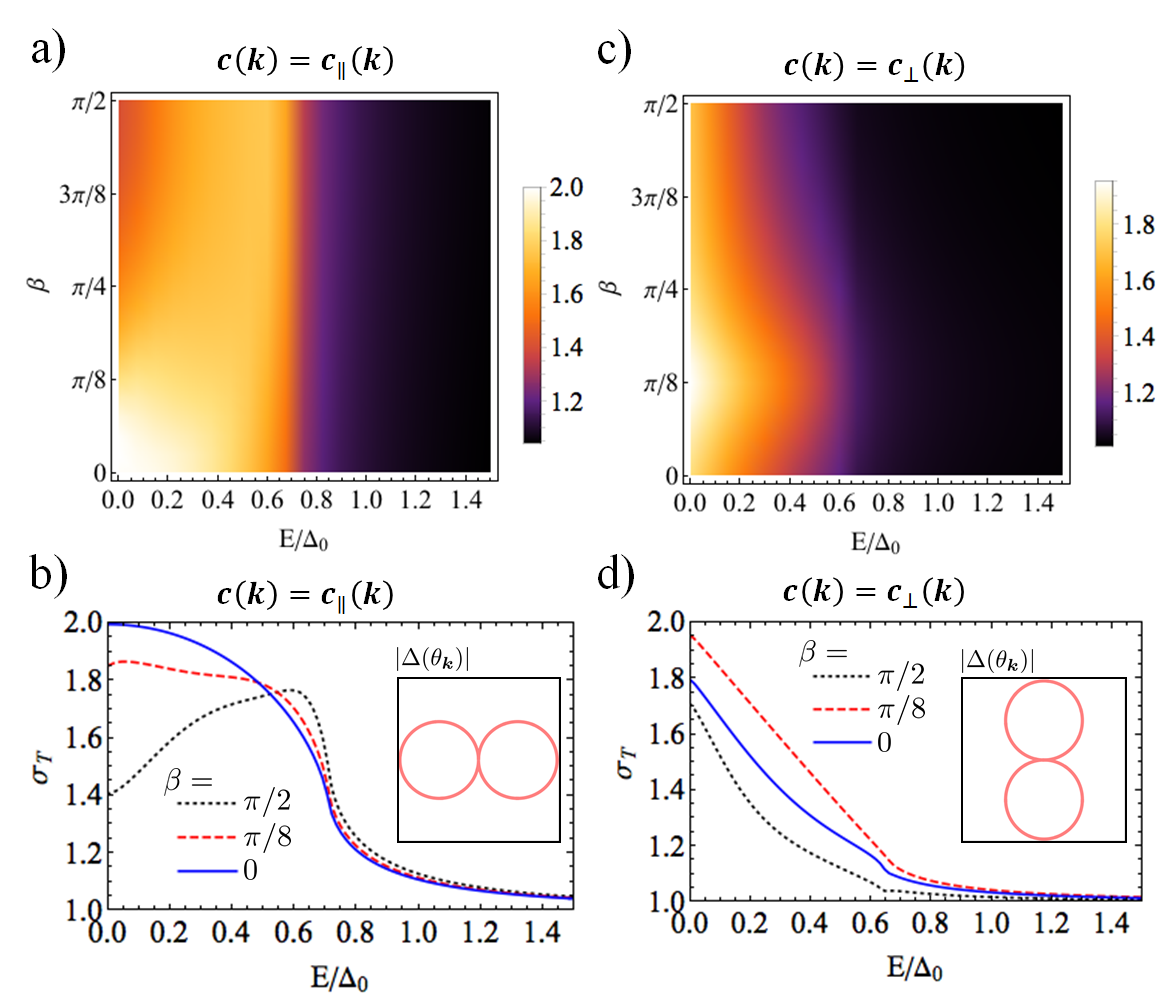}
\caption{Conductance of a lateral heterojunction  with a chiral parent SC. Panels a) and b) (panels c) and d)) are for the TI-SC interface described by $\bm{c}_{\parallel}(\bm{k})$ ($\bm{c}_{\perp}(\bm{k})$), see text. Top row shows the color map as a function of the boundary parameter $\beta$ in Eq.~\eqref{matrixm}, bottom row shows 
$\sigma_{T}$  for select values of $\beta$. The insets give the shape of the induced superconducting gap at the interface.}
\label{chiral}
\end{figure}

\textit{Conductance of Lateral Heterojunctions.}  Directly measuring the gap in the interface layer of a TI-SC heterostructure is technically difficult. Therefore we consider the signatures of the induced order parameter in the conductance of lateral heterojunctions, shown in Fig.~\ref{fig:mags}(b), between region I with the usual vacuum-terminated TI surface defined by $\bm c(\bm k)=\bm c_{\rm D}(\bm k)$ and region II with the TI-SC interface. To explore the salient consequences of the rotational symmetry breaking in the interface, we consider two simplified generic choices for the interface states, namely $\bm{c}_{\parallel}(\bm{k}) = \hbar v_{\rm F}(k_{y},-\lambda k_{x},-\lambda k_{x})$ and  $\bm{c}_{\perp}(\bm{k}) = \hbar v_{\rm F}(\lambda k_{y},-k_{x},-\lambda k_{y})$. In the following we present the results for $\lambda = 2/3$ chosen such that $\bm{c}_{\parallel}(\bm{k}) = \bm{c}_{u_{\text{max}}}(\bm{k})$, so that the major axis of the elliptical Fermi surface for $\bm c_\|(\bm{k})$ ($\bm c_\perp(\bm{k})$) is parallel to the $x$ ($y$) axis. Crucially, with these choices the gap nodes of the proximity-induced superconducting order parameter from the chiral parent SC are along the $k_x=0$ direction for $\bm c_\|(\bm{k})$ and along the $k_y=0$ direction for $\bm c_\perp(\bm{k})$, as shown in the insets of Figs.~\ref{chiral}(b) and (d). 

This difference is manifested in the conductance due to Andreev reflection at the lateral junction. We compute the conductance spectra using the semi-classical Andreev equations~\cite{Snelder2015,Blonder1982,Bruder1990,Kashiwaya1996}. Correct boundary conditions at the junction are critical for the wavefunction matching in the Andreev approach. These boundary conditions are not trivial for effective Hamiltonians that are linear in momentum, as is well-known from studies of graphene edge states~\cite{McCann2004,Basko2009,Akhmerov2008}. We model the lateral edge~\cite{Asmar2018a} by setting $\Psi_{\rm I}(0) = \mathcal{M}\Psi_{\rm II}(0)$. The matrix $\mathcal{M}$ is determined from the requirement that the Hamiltonian for the lateral junction is Hermitian, i.e. conserves the probability current. If in addition the heterostructure preserves time-reversal symmetry, we find a one-parameter family of boundary conditions described by~\cite{Asmar2018a}
\begin{equation}
\mathcal{M}(\beta) = \sqrt{\dfrac{v}{v_{\rm F}}}\bigg[\tau_{0} e^{i\sigma_{y}\beta} + \dfrac{i\tau_{z}}{2\hbar v} (c_{xx}\sigma_{z} - c_{zx}\sigma_{x})e^{-i\sigma_{y}\beta}\bigg].
\label{matrixm}
\end{equation} 
Here $\tau_{i}$ are the Pauli matrices in the electron-hole space as written in the basis of Eq. \eqref{bdg}, and $v = (\sqrt{\sum_{i}c_{ix}^{2}} - c_{yx})/2\hbar$. In the spirit of the weak-coupling limit we derived ${\cal M}(\beta)$ in the normal state, and neglected any modification of it due to the emergence of proximity-induced superconductivity. The parameter $\beta$ thus encodes all possible scattering phenomena at the junction compatible with time-reversal symmetry and particle conservation, and accounts for the fact that we are working with an effective low energy Hamiltonian~\cite{Ahari2016,Isaev2015}.

In the Andreev approximation we solve for wavefunctions of the form $\Phi(\bm{r}) = e^{i\bm{k}_{\rm F}\cdot \bm{r}}(U(\bm{r}),0,V(\bm{r}),0)$, as given in the basis of Eq. \eqref{ham}, where $\bm{k}_{\rm F}$ is the Fermi momentum. From the BdG equation $\mathcal{H}\Phi(\bm{r}) = E\Phi(\bm{r})$, we obtain the Andreev equations for the envelope functions $U(\bm{r})$ and $V(\bm{r})$,
\begin{equation}
\begin{aligned}
E U(\bm{r}) = -i\bm{v}_{\rm F}(\bm{k}_{\rm F})\cdot \bm{\nabla} U(\bm{r}) + \Delta(\bm{r},\bm{k}_{\rm F})V(\bm{r}),
\\ E V(\bm{r}) = i \bm{v}_{\rm F}(\bm{k}_{\rm F})\cdot \bm{\nabla} V(\bm{r}) + \Delta^{*}(\bm{r},\bm{k}_{\rm F})U(\bm{r}).
\end{aligned}
\end{equation}
Here $\bm{v}_{\rm F}(\bm{k}_{\rm F}) = (\partial\xi/\partial\bm{k})|_{\bm{k}_{\rm F}}$ with $\xi(\bm{k}) = |\bm{c}(\bm{k})| - \mu$ being the normal state energy dependent on whether we are in region I or II. We approximate $\Delta(\bm{r},\bm{k}_{\rm F}) = \Delta(\bm{k}_{\rm F})\Theta(x)$, where $\Delta(\bm{k}_{\rm F})$ is given by Eq. \eqref{op} and $\Theta(x)$ is the Heaviside step function. Upon obtaining the wavefunctions, we transform back to the basis of Eq. \eqref{bdg} and match the solutions using the matrix $\mathcal{M}(\beta)$ as given by Eq. \eqref{matrixm}. We then consider an incoming incident electron in region I with momentum $\bm{k}_{1} = (k_{x},k_{y})$ and in-plane angle $\theta_{\bm{k}}$. At the boundary, the electron may be retroreflected as a hole with momentum $\bm{k}_{1}$, or specularly reflected as an electron with momentum $\bm{k}_{2} = (-k_{x},k_{y})$. The wavefunction in region I is then $\Psi_{\rm I}(\bm{r}) = \Psi_{\text{e}}^{\text{inc}}(\bm{r}) + A\Psi_{\text{h}}^{\text{ref}}(\bm{r}) + B\Psi_{\text{e}}^{\text{ref}}(\bm{r})$. In region II, the incident electron can either be transmitted as an electron-like quasiparticle with momentum $\bm{k}'_{1} = (k'_{x},k_{y})$, or a hole-like quasiparticle with momentum $\bm{k}'_{2} = (-k'_{x},k_{y})$.

We then solve for the coefficients $A$ and $B$. For a fixed energy $E$, the transmission coefficient for electrical current\cite{Blonder1982,Tanaka1995,Kashiwaya1996} can be defined as $\sigma_{S} (E,\theta_{\bm{k}}, \beta)= 1 + |A|^{2} - |B|^{2}$. Normalizing to the normal state value, $\sigma_N$, the total dimensionless conductance is defined as\cite{Tanaka1995,Kashiwaya1996}
\begin{equation}
\sigma_{T}(E,\beta) = \dfrac{\int_{-\pi/2}^{\pi/2}\sigma_{S}(E,\beta,\theta_{\bm{k}})\cos\theta_{\bm{k}} d\theta_{\bm{k}}}{\int_{-\pi/2}^{\pi/2}\sigma_{N}(\beta,\theta_{\bm{k}})\cos\theta_{\bm{k}} d\theta_{\bm{k}}}.
\label{eq:sigma}
\end{equation}
As may be expected, quasiparticles at near normal incidence give a dominant contribution to the conductance. For our choice of $\bm c_\|(\bm k)$ these quasiparticles see the full gap, while for $\bm c_\perp(\bm k)$ such quasiparticles travel along near-nodal directions. Consequently, we expect the features associated with superconductivity to be much more prominent for the former case.

Fig.~\ref{chiral} shows precisely this behavior. The conductance for the $\bm c_\perp(\bm{k})$ case rapidly, and almost linearly, decreases from the value near $\sigma_T\approx 2$ characteristic of Andreev reflection, while for $\bm c_\|(\bm{k})$ the conductance retains a more typical shape with a decrease from $\sigma_T = 2$ to $\sigma_T\to 1$ at energies comparable with the amplitude of the interface superconducting gap. Notably, this feature is robust with respect to the variations of the boundary parameter $\beta$, and therefore provides an unmistakable experimentally accessible signature of the symmetry breaking at the TI-SC interfaces.

\textit{Discussion}. There are two parts to our analysis. First, we derived the form of the proximity-induced order parameter for an arbitrary parent superconductor and any set of coefficients $c_{ij}$ characterizing the topological interface state, Eq.~\eqref{op}, and showed that symmetry breaking at interfaces may drastically alter superconductivity in the interface layer.  The most dramatic changes relative to previously studied cases occur for triplet systems, where the existence and shape of the proximity-induced gap are controlled by the deviations from the helical Dirac surface spectrum, $\bm c_{\rm D}(\bm k)$. The second part of our analysis focused on the particular case of the proximity effect with a fully gapped chiral triplet superconductor, and found that the superconductivity in the interface layers exists whenever $c_z(\bm k)\neq 0$, and is nodal. We demonstrated that Andreev spectroscopy of the lateral heterojunctions provides information on the nodal structure of the proximity-induced gap, and therefore tests for the presence of such symmetry breaking interface potentials. We also found that the qualitative structure of the spectra is independent of the details of the scattering at the lateral junction. 

While some of our results were obtained within a specific model of the interface, they are based on symmetry considerations and therefore we expect them to remain qualitatively correct irrespective of the detailed origin of the symmetry breaking. Our classification of the superconducting chiral states was done for the tetragonal D$_{4h}$ symmetry, while the $[111]$ (in the rhombohedral unit cell) plane of the prototypical Bi$_2$Se$_3$ and related topological insulators has the hexagonal D$_{6h}$ symmetry. To lowest order in $\bm k$, however, this does not change the results for the chiral order parameter and therefore our results hold. Away from the Dirac point, higher-order momentum corrections due to hexagonal warping may result in the appearance of out-of-plane spin components of the interface states~\cite{Fu2009}, but the corresponding contribution to the order parameter is smaller than the leading order effect discussed here, and leads to a subdominant 6-fold modulation of the superconducting order parameter.   In a broader context our results pave the way for the targeted design of superconducting proximity-induced orders via interface engineering. 

\textit{Acknowledgments}. This research was supported by NSF via Grants No. DMR-1410741 (D.J.A., M.M.A. and I.V.) and No. DMR-1151717 (M.M.A. and D.E.S.).

\end{document}